\documentclass[a4paper]{jpconf}
\usepackage{tablefootnote}
\usepackage{lmodern}
\usepackage{amssymb,amsfonts,amsmath,,amscd}
\usepackage{subeqnarray}
\usepackage{extarrows}
\usepackage[usenames,dvipsnames]{color}
\usepackage{stmaryrd}
\usepackage{amsmath,epsfig}
\usepackage{amscd,amsbsy,array}
\usepackage{dsfont}

\begin{document}
\title{KAPPA-DEFORMATIONS:  HISTORICAL DEVELOPMENTS  AND RECENT RESULTS}

\author{Jerzy Lukierski\footnote{Presented at ISQS24 Conference on Integrable Systems and Quantum Symmetries, Prague (Czech Republik), 14.06--18.06.2016, 3-rd POTOR (Polish Society on Relativity) Conference, Cracov (Poland), 25--29.09.2016 and 5-th Conference ``New Trends in Field Theories'', Varanasi (India), 06.11--10.11.2016.}}

\address{Institute for Theoretical Physics, University of Wroc\l aw (Poland)}

\ead{jerzy.lukierski@ift.uni.wroc.pl}

\begin{abstract}
 I shall recall  in historical perspective  
 some results from nineties and show further how $\kappa$-deformed symmetries and $\kappa$-Minkowski space inspired DSR (Doubly of Deformed Special Relativity) approach proposed after 2000. As very recent development I shall show how to describe quantum-covariant $\kappa$-deformed phase spaces by passing from Hopf algebras to Hopf algebroids (arXiv:1507.02612) and I will briefly
describe the  $\kappa$-deformations of  $AdS_5 \times S^5$ superstring target spaces (arXiv:1510.030.83).
\end{abstract}

\section{Introduction}

Transition from classical to quantum physics leads to the appearance of noncommutative algebraic structures. In standard quantum mechanics (QM) the canonical quantum phase space is described by Heisenberg algebra
 (HA) (i,j=1,2,3)
\begin{subeqnarray}\label{jlprag1}
&\hbox{HA in}
\qquad
&\left[ \hat{x}_i, \hat{x}_j \right] = \left[ \hat{p}_i , \hat{p}_j \right] = 0
\label{jlprag1a}
\\
&\hbox{standard QM}  \qquad
& \left[ \hat{x}_i, \hat{p}_j \right] = i\hbar \, \delta_{ij}
\label{jlprag1b}
\end{subeqnarray}

The noncommutativity (\ref{jlprag1b}b) of positions $\hat{x}_i$ and momenta $\hat{p}_i$  can be used to derive the Heisenberg uncertainty relations which introduce the bounds on accuracy of simultaneous measurements of positions and momenta.
Besides it follows from (\ref{jlprag1a}a) that in standard QM one can localize separately the positions
or momenta with arbitrary accuracy, what is reflected in the use in QM of classical geometry, with commutative space and time.

The canonical nonrelativistic HA (\ref{jlprag1a}a--\ref{jlprag1b}b)  is changed  however if 
$(\hat{x}_i,\hat{p}_i)$ describe a dynamical system, e.g. point particles  moving in a field-theoretic
background.
In particular, in the presence of electromagnetic (EM) fields the canonical momenta are becoming noncommutative
(NC).
Nonrelativistic  EM background described by magnetic field $\vec{H}= \textrm{rot} \vec{A}$
($H_i = \frac{1}{2} \varepsilon_{ijk}F_{jk}$  where $F_{jk}=\partial_j A_k - \partial_k A_j$), leads to the following modification of algebra (\ref{jlprag1a}a--\ref{jlprag1b}b)  \cite{jlrprag1,jlrprag2}
\begin{subeqnarray}\label{jlprag2}
&\hbox{HA in the presence}
\qquad
&\left[ \hat{x}_i, \hat{x}_j \right] = 0  \qquad \left[ \hat{p}_i , \hat{p}_j \right] = i e\, F_{ij}
\label{jlprag12a}
\\
&\hbox{of magnetic field H}  \qquad
& \left[ \hat{x}_i, \hat{p}_j \right] = i\hbar \, \delta_{ij}
\label{jlprag2b}
\end{subeqnarray}

In relativistic theories one uses the Poincar\'{e} algebra as describing the group of motions in Minkowski space-time.
The presence of a constant relativistic EM background ($F_{\mu\nu}=\partial_\mu A_\nu - \partial_\nu A_\mu$; $\mu,\nu=0,1,2,3$) leads to the deformation of  Poincar\'{e} algebra into Maxwell algebra \cite{jlrprag3}--\cite{jlrprag5}, with noncommutative fourmomenta generators $P_\mu$

\begin{equation}\label{jlprag3}
\begin{array}{c}
\left[ P_\mu, P_\nu \right]= 0
\\
\hbox{(Poincar\'{e} algebra)}
\end{array}
\qquad 
{\xlongrightarrow{F_{\mu\nu}\neq0}} 
\qquad
\begin{array}{c}
\left[ P_\mu , P_\nu \right] = i  Z_{\mu\nu}
\\
\hbox{(Maxwell algebra)}
\end{array}
\end{equation}
where new six Abelian generators $Z_{\mu\nu} = e F_{\mu\nu}$ describe  the tensorial central charges.

Further let us consider quantum-mechanical system in the presence of gravitational field ($g_{\mu\nu}(x)$
 or $e^a_{\ \mu}(x)$, where $g_{\mu\nu}= e^{\ a}_\mu \, \eta_{ab} \, e^{\ b}_\nu$;  $\eta_{ab}= \textrm{diag}(-1,1,1,1)$). 
 It has been argued   \cite{jlrprag6,jlrprag7} that one gets in quantum regime for such a system  the deformed 
canonical phase space with noncommutative positions sector,  because the quantum fluctuations of gravitational field does not allow
the localization of positions (space-time coordinates) with arbitrary accuracy.
Such restriction of measurements has been physically interpreted in the framework of  Einstein  gravity as caused by the formation at Planckian distances ($\lambda_p \simeq \textrm{10}^{-33}\textrm{cm}$) of mini-black holes, screening experimental visibility of  subPlanckian distances ($\lambda < \lambda_p$)\footnote{Such mechanism was firstly qualitatively predicted by Bronstein in 1936 \cite{jlrprag8}.  He predicted that the geometry describing all interactions should incorporate three 
 fundamental constants: $c,\hbar$ and Newton constant $G$; in place of  $G$  there was later introduced the  Planck mass $m_{\rm pl}=(\frac{h c}{G})^{1/2} \simeq 10^{-5}g$.}.

Summarizing, if we consider  quantized dynamical systems interacting with  gravity (QG) one should use new type 
of quantum phase spaces, with noncommuting  coordinates, in relativistic quantum theories described by deformed noncommutative Minkowski spaces.  We get
\begin{equation}\label{jlprag4}
\begin{array}{c}
\left[ \hat{x}_\mu , \hat{x}_\nu \right] = 0
\\
\hbox{space-time without}
\\
\hbox{QG effects}
\end{array}
\qquad
\stackrel{\hbox{nonvanishing}}{\overrightarrow{\  \  \  \hbox{QG effects} \  \  }}
\qquad
\begin{array}{c}
\left[ \hat{x}_\mu, \hat{x}_\nu \right] \neq 0
\\
\hbox{at Planckian}
\\
\hbox{distances}
\end{array}
\end{equation}
One can point out that noncommutative (quantum) space-time coordinates represent {\color{blue}new geometric
 paradigma in theoretical physics},   {\color{blue}providing the description of quantum dynamical systems with   inclusion of quantum gravity effects}.

Let me report briefly on the appearance of noncommutative geometry, quantum spaces and particularly quantum space-times in mathematical physics.
First noncanonical noncommutative structures were introduced around 1980 as so-called quantum algebras describing the  algebraic and geometric properties of quantum integrable system \cite{jlrprag9}--\cite{jlrprag11}.
These algebro-geometric structures were introduced as one-parametric deformations of  universal enveloping algebra for some low-dimensional complex Lie
 algebras, and  supplemented with additional Hopf-algebraic structure, i.e. with coalgebras (coproducts), 
coinverses (antipodes) and counits\footnote{For Hopf-algebraic description of deformed groups and Lie
algebras see e.g. \cite{jlrprag12}--\cite{jlrprag14}.}.
The deformed quantum Hopf algebras  were proposed as noncommutative  and noncocommutative generalizations
of the notions  of classical Lie groups  $G$ and Lie algebras  $\hat{g}$.
First classification of deformed simple complex Lie algebras $\hat{g}$ (the Hopf deformations of enveloping Lie algebras $U(\hat{g})$) was given by Drinfeld \cite{jlrprag12} and Jimbo \cite{jlrprag15}, who introduced standard quantum deformations,  called as well  Drinfeld-Jimbo (DJ) or $q$-deformation ($q$ - complex deformation parameter).
In eighties, in parallel  way, there were also introduced the quantum groups as Hopf-algebraic deformations of matrix Lie groups  $G$,  described by functions on matrix groups with suitably introduced    noncommutative matrix entries \cite{jlrprag16,jlrprag17,jlrprag13}.
Subsequently, using Hopf-algebraic duality, one gets the description of quantum symmetries by pairs of dual
 Hopf algebras generalizing dual  pairs of classical matrix Lie groups and classical Lie algebras, linked by  exponential map (passage
from Lie algebras to Lie group) or differentiation of Lie group elements at group unit (passage from Lie groups to Lie algebras). 

Quantum symmetries are described in their global (finite) and infinitesimal versions as follows:

\begin{equation}\label{jlprag5}
\begin{array}{c}
\hbox{infinitesimal}
\\ 
\hbox{quantum symmetries}
\\
\hbox{(quantum Lie}
\\
\hbox{algebras)}
\end{array}
\qquad
\stackrel{\hbox{Hopf-algebraic}}{\overleftrightarrow{ \  \  \  \  \  \  \hbox{duality} \  \  \  \  \   \  }}
\qquad
\begin{array}{c}
\hbox{finite  }
\\ 
\hbox{quantum symmetries}
\\
\hbox{(quantum matrix}
\\
\hbox{groups)}
\end{array}
\end{equation}

\noindent
We add that quantum matrix groups, with noncommutative group elements, were also introduced  as the linear transformation groups  describing the quantum symmetries  of various noncommutative spaces (see e.g. \cite{jlrprag17}).

The basic problem at the end of eighties was the consistent quantum deformation of relativistic symmetries:
$D=4$ Lorentz and Poincar\'{e} algebras.
The explicit $q$-deformation of  Lorentz algebra was easily introduced as particular application of general DJ deformation framework \cite{jlrprag18}--\cite{jlrprag21}.
However, to obtain quantum deformation of non-semisimple Poincar\'{e} algebra a new way of deriving 
Hopf-algebraic quantum deformations was needed\footnote{Few years later it was shown that DJ $q$-deformation of  D=4 Lorentz algebra can be as well extended to \hbox{$q$-deformed} D=4 Poincar\'{e} algebra, but in the framework of  nonstandard braided Hopf algebras \cite{jlrprag21}, with coalgebras defined with the use of nonstandard braided tensor product.}. 

The quantum deformations of important class of nonsemisimple classical Lie algebras, described by semidirect product of Abelian and simple Lie subalgebras, were obtained by the quantum modification of known Wigner-Inonu (WI) contraction procedure \cite{jlrprag22,jlrprag23}. 
Firstly such quantum WI contraction was applied to the derivation of  quantum $\kappa$-deformation $U_\kappa(e(2))$ of D=2
Euclidean inhomogeneous algebra $e(2) = O(2) \ltimes T^2 $ \cite{jlrprag24}
\begin{equation}\label{jlprag6}
U_q(su(2)) \qquad 
\begin{array}{c}
\phantom{xx}
\\[-14pt]
\stackrel{\hbox{$q(R)$}}{\overrightarrow{ \  \  \    { R \to \infty} \  \  \  }}
\end{array}
\qquad U_\kappa(e(2))
\end{equation}
where $R$ denotes the WI contraction parameter,  $U_q(su(2))$ describes DJ deformation of enveloping algebra $U(su(2))$ 
 and 
{\hbox{\footnotesize{$\stackrel{{q(R)}}{
{\hbox{\tiny{${\overrightarrow{\  \  \  R\to \infty \  \ }}$}}}
}$}}}
denotes the quantum WI  contraction limit $R\to \infty$ with $q(R) \to 1$,
  which requires the special choice of $R$-dependence of $q(R)$,
 provided by the following asymptotic expansion

\begin{equation}\label{jlprag7}
q(R) = 1 + \frac{1}{\kappa R} +  O\left(  \frac{1}{(\kappa R)^2}\right)
\  \  
\begin{array}{c}
\phantom{xx}
\\
 {\overrightarrow{ \  \  \  R \to \infty \  \  \ }}  
\end{array}
\   1
\end{equation}

The particular $R$-dependence given by  (\ref{jlprag7}) leads to finite but nonclassical (deformed) contraction limit, defining $\kappa$-deformed Hopf  algebra $U_\kappa(e(2))$.
Because for space-time symmetries $R$ has the length dimension $\left[L\right]$ and  $q(R)$ should be dimensionless,  it follows that  in  (\ref{jlprag7})  the parameter $\kappa$ has the dimensionality $\left[ L^{-1}\right]$, what  (after putting $c=\hbar=1$)  implies the mass dimensionality of $\kappa$.
In such a way there were introduced quantum-deformed  D=2 inhomogeneous Euclidean symmetries with mass-like deformation parameter $\kappa$, which has been further  identified with the Planck mass $m_{\rm pl}$.

Our presentation reviews briefly three periods of research activity.
\\[2pt]
\noindent
i) last decade of XXth century, when the description of $\kappa$-deformed quantum Hopf-algebraic symmetries was  proposed and well established, 
\\[2pt]
\noindent
ii) first decade of present century, with the development of  rather semi-phenomenological  approach to the deformation of special relativity, called    Doubly Special Relativity  or  later Deformed Special Relativity (DSR),  
\\[2pt]
\noindent
iii) recent years, when  several new applications of $\kappa$-deformations were  proposed.

The plan of our presentation is the following:

In Sect.~\!\!2 \,  I recall the introduction in 1991--92 of  $\kappa$-deformed D=4 Poincar\'{e} algebra \cite{jlrprag25}--\cite{jlrprag27}, obtained by  quantum Wigner-Inonu contraction procedure (see (\ref{jlprag6}--\ref{jlprag7}))  applied to suitable Lie algebras of rank 2.
Further  it is presented  briefly the problem of  choice of algebraic basis for $\kappa$-deformed Poincar\'{e}-Hopf algebra, and  we   provide two important explicit  choices: Majid-Ruegg bicrossproduct basis \cite{jlrprag28,jlrprag29} 
  and  classical Poincar\'{e} algebra basis \cite{jlrprag30}--\cite{jlrprag32}.

In Sect.~\!\!3\,  I describe the notion of duality of Hopf algebras and define $\kappa$-deformed  Poincar\'{e}-Hopf group as dually determined by the  $\kappa$-deformed Poincar\'{e}-Hopf algebra (see e.g. \cite{jlrprag33,jlrprag34}).
It will be shown  how by using $\kappa$-deformed Poincar\'{e} algebra  in a selected basis one can obtain the finite $\kappa$-deformed quantum Poincar\'{e} group transformations.

In Sect.~\!\!4 \, I consider  briefly the postulates of DSR theories \cite{jlrprag35}--\cite{jlrprag37} and expose their  link with earlier descriptions  of  $\kappa$-deformed Poincar\'{e} symmetries.
One can show that basic two  formulae of DSR theories providing the deformation of mass-shell condition and quantum deformation of finite Lorentz transformations  were  already present in earlier Hopf algebraic description of  $\kappa$-deformed quantum symmetries (see e.g. \cite{jlrprag38}).

Further I shall recall subsequent  development of DSR theories, with coalgebraic part of 
  \hbox{$\kappa$-deformed} formalism represented as curved structures of classical momentum spaces \cite{jlrprag39}--\cite{jlrprag42}. 
 I~will  comment as well on other deformations which introduce mass-like deformation parameter, in particular
 the oldest one, the Snyder deformation of space-time \cite{jlrprag43}, which introduces mass-like deformation parameter 
 (e.g. Planck mass)  without violating the classical Lorentz symmetry.

In Sect.~\!\!5 \,  I  shall describe briefly two recent applications of $\kappa$-deformations: the introduction of quantum-covariant $\kappa$-deformed phase space as an example of quantum space with Hopf algebroid structure \cite{jlrprag44,jlrprag45} and the insertion of  $\kappa$-deformation into the Yang-Baxter (YB) sigma model defined on
 $\frac{SU(2,2;4)}{O(4,1)\times O(5)}$ coset space, which leads to the $\kappa$-deformation of D=10 GS superstring target space geometry \cite{jlrprag46}.

Finally, in Sect.~\!\!6,   an outlook is presented. 
The $\kappa$-deformation is an example of  a~global, nondynamical deformation of space-time algebra, which  only approximates  ``physical''  noncommutative structures in QG.
More general deformations, local and of dynamical  origin, should relate the  parametrization of noncommutativity with dynamical degrees  of freedom  of   quantum gravity + quantum matter systems.
New formalism determining  such relations will provide  in future the final fundamental quantum theory, describing interacting QG with the quantized matter fields.
\section{$\kappa$-deformed Poincar\'{e}-Hopf algebras}
Drinfeld-Jimbo (DJ) $q$-deformation scheme \cite{jlrprag12} has been proposed only for complex simple Lie
algebras.
Because space-time groups of motions are real and nonsemisimple, in order to obtain e.g. quantum-deformed D=4 Poincar\'{e} algebra $\hat{p}_{3,1}$, which is a nonsemisimple cross product of real Lorentz and translation (Abelian) subalgebras ($\hat{p}_{3,1} = \hat{o}(3,1) \supsetplus T_4$), one should apply the quantum WI contraction procedure.
In four dimensions in place of (\ref{jlprag6}) we perform the contraction\footnote{The contraction (\ref{jlprag8}) provides undeformed $\hat{o}(3)$ rotation. 
If we apply quantum WI contraction to $U_q(o(4,1))$ we obtain $\kappa$-deformation of D=4 Poincar\'{e} algebra with undeformed $\hat{o}(2,1)$ subalgebra (see \cite{jlrprag47}, table~1.)}
\begin{equation}\label{jlprag8}
U_q(\hat{o}(3,2))  \quad   
\begin{array}{c}
\phantom{xxx}
\\[-10pt]
\stackrel{\hbox{$q(R)$}}{\overrightarrow{ \  \  \    { R \to \infty} \  \  \  }}  
\end{array}
\quad 
U_\kappa(\hat{p}_{3,1})
\end{equation}
where $q(R)$ has asymptotic expansion (\ref{jlprag7}).

To perform explicitly  the contraction (\ref{jlprag8}) we should extend  DJ deformation of Cartan-Chevaley
 (CC) basis of   complexified $\hat{o}(3,2)$ algebra $(Sp(4;c)\simeq o(5;c))$,  defined for  6 generators $h_i, e_{\pm i}$,  ($i=1,2$),  to Cartan-Weyl (CW) Lie-algebraic basis
 (10 generators extending $\hat{o}(3,2)$ CC basis by four generators $e_{\pm 3}, e_{\pm 4}$  \cite{jlrprag47,jlrprag48}).
  We resolve in such a way the $q$-deformed Serre relations formulated in DJ deformation scheme
\footnote{We point out that $q$-deformations of Cartan-Weyl bases for rank two Lie algebras and superalgebras has been firstly presented in explicite form in \cite{jlrprag48}.}.
Further, because the D=4 physical space-time symmetry algebras ($\hat{p}_{3,1}, \hat{o}(3,2), \hat{o}(4,1), \hat{o}(4,2)$ etc.) are real, with Hermitean or antiHermitean generators, one should study all possible reality conditions, and consider corresponding real  \hbox{$\star$-Hopf} algebras.
There were used in literature mainly the 
following two types of   real $\star$-Hopf algebras:
\\[4pt]
\indent
i) Standard $\star$-Hopf algebras, with  involutive anticonjugation
 $(x,y)^\star = y^\star x^\star$, which maps  tensor products in nonflipped way $((x\otimes y)^\star = x^\star \otimes y^\star) $ \cite{jlrprag27,jlrprag28}.
Such real $q$-deformed $\star$-Hopf algebra $U_q(\hat{o}(3,2))$ with $q$ real was used to obtain by quantum WI contraction the $\kappa$-deformed Poincar\'{e} algebra as standard real $\star$-Hopf algebra (for notion of  standard $\star$-Hopf algebras see e.g. \cite{jlrprag49}, \cite{jlrprag50}).
\\[4pt]
\indent
ii) nonstandard $\star$-Hopf algebra, with involutive conjugation $(xy)^\star = x^\star \cdot y^\star$ and flipped 
conjugation of tensor products $(x\otimes y)^\star = y^\star \otimes x^\star$, which is for example  used in $q$-deformed models with 
 $q$ described by roots of unity \cite{jlrprag51,jlrprag52}.
 Such nonstandard reality conditions was also used in first derivations of  $\kappa$-deformed Poincar\'{e} algebra
 \cite{jlrprag25,jlrprag26} obtained from $q$-deformed AdS algebra $\hat{o}(3,2)$ with $|q|=1$.

The quantum WI contraction should be performed on all Hopf-algebraic operations, i.e.

\begin{equation}\label{jlprag9}
\mathds{H}_q = (A_q= U_q(\hat{o}(3,2)), m, \Delta_q, \epsilon, S_q)  \  \
\begin{array}{c}
\phantom{xx}
\\[-14pt]
{\hbox{\small{$
 \stackrel{\hbox{$q(R)$}}{\overrightarrow{ \  \  \    { R \to \infty} \  \  \  }}
$}}}
\end{array}
 \ \  \mathds{H}_\kappa= (A_\kappa = U_\kappa(\hat{p}_{3,1}), m, \Delta_\kappa, \epsilon, S_\kappa)
\end{equation}
where $m: A\otimes A \to A$ denotes multiplication, $\Delta: A \to A\otimes A$ the coproduct and $S: A\to A$
 the antipode (coinverse).

The $\kappa$-deformation of Poincar\'{e} algebra  obtained by (\ref{jlprag9}) contains undeformed $\hat{o}(3) \supsetplus \hat{T}_3$
 subalgebra describing nonrelativistic D=3 Euclidean group of motions.

Using three-dimensional notation for D=4 Lorentz generators $M_{\mu\nu}=(M_i, N_i)$ (i=1,2,3) and 
fourmomenta $P_\mu = (P_i, P_0)$,  there were obtained in   \cite{jlrprag25}--\cite{jlrprag27} 
  the following deformations

\begin{description}
\item[{--}\ ]  modified commutators $\left[ N_i , N_j \right]$ and $\left[ P_i , N_j \right]$

\item[{--}\ ]  nonprimitive coproducts $\Delta(N_i)$, $\Delta(P_i)$

\item[{--}\ ]  modified antipodes (in undeformed case $S(P_\mu)=-P_\mu, \, S(M_{\mu\nu})=-M_{\mu\nu}$)
\end{description}

\noindent
Subsequently it was realized that the suitable change of linear Poincar\'{e} algebra generators
\begin{equation}\label{jlprag10}
M_{\mu\nu} \to M'_{\mu\nu}(M_{\mu\nu}, P_\mu)
\qquad \qquad
P_\mu \to P'_\mu = P'_\mu(P_\mu)
\end{equation}
can shift some deformations from algebra to coalgebra.
The following two bases became very useful and subsequently used:
\\[6pt]
\noindent
\textit{i) Majid-Ruegg basis \cite{jlrprag28}}
\\[3pt]
\indent
In such a basis the $\kappa$-Poincar\'{e}-Hopf algebra is described as follows:
\\[4pt] 
a) Algebra 
\begin{description}
\item[{--}\ ] Lorentz algebra remains classical, i.e. described by $o(3,1)$ Lie algebra

\item[{--}\ ]  Space-time translations  generators $P_\mu$ commute (i.e. also remain classical)

\item[{--}\ ]  The only  $\kappa$-deformed Poincar\'{e} algebra commutator looks as follows:
\end{description}
\begin{equation}\label{jlprag11}
\left[ N_i , P_j \right] = i \delta_{ij} \left[ 
\frac{\kappa}{2} (1- e^{- \frac{2P_0}{\kappa}}) + 
\frac{1}{2\kappa^2} \vec{P}^2
\right] + \frac{1}{\kappa} P_i P_j 
\end{equation}

\noindent
b) Coalgebra 

The coproducts $\Delta(P_0)$ and $\Delta(M_i)$  remained primitive (classical), the  ones which were  deformed  consistently with
 (\ref{jlprag11}) are the following
\begin{eqnarray}\label{jlprag12}
&& \Delta P_i = P_i \otimes 1 + e^{- \frac{P_0}{\kappa}} \otimes P_i
\nonumber \\  \\[-10pt]
&& \Delta N_i = N_i \otimes 1 + e^{- \frac{P_0}{\kappa}} \otimes N_i + \frac{1}{\kappa} \in_{ijk} P_j \otimes M_k
\nonumber
\end{eqnarray}
\\[4pt]
\noindent
c) Antipodes (coinverses)
\begin{eqnarray}\label{jlprag13x}
&& S(M_i) = - M_i   \qquad \qquad  S(N_i) = - N_i + \frac{3i}{\kappa} P_i
\nonumber  \\  \\[-10pt]
&& S(P_i) = - e^{-\frac{P_0}{\kappa}} P_i  \qquad S(P_0)= - P_0
\nonumber
\end{eqnarray}

\noindent
d) Casimir operators

d1) The $\kappa$-deformed mass Casimir,  defining $\kappa$-deformed mass-shell condition, takes the form
\begin{equation}\label{jlprag13}
C_2 = \vec{P}^2  \, e^{\frac{P_0}{\kappa}} - (2 \kappa \sin \frac{P_0}{\kappa})^2
\end{equation}

d2) The $\kappa$-deformed spin-square Casimir looks as follows 
\begin{equation}\label{jlprag14}
C_4 =  ( \cosh  \frac{P_0}{\kappa} - \frac{\vec{P}^2}{4\kappa^2})W^2_0 - \vec{W}^2_\kappa
\end{equation}
where the component $W_0$ is undeformed\footnote{The fourvector $W_\mu= (W_0, W_i)$ was
introduced in undeformed case by Lubanski \cite{jlrprag53}. It is called Pauli-Lubanski fourvector.}
\begin{equation}\label{jlprag15}
W_0 = \vec{P} \, \vec{M}
\end{equation}
but three-vector  $\vec{W}_\kappa$  is $\kappa$-dependent
\begin{equation}\label{jlprag16}
\vec{W}_{\kappa} = \kappa \vec{M} \,
 \sinh \frac{P_0}{\kappa} + \vec{P}\times \vec{N}
\end{equation}
\\
\textit{ii) classical Poincar\'{e} basis \cite{jlrprag30}--\cite{jlrprag32}}

In such a basis the whole Poincar\'{e} algebra is not deformed, and all deformations are incorporated into quite
complicated coproducts (in particular the coproduct for energy generator  $\Delta(P_0)$  becomes  nonprimitive).

The $\kappa$-deformed Poincar\'{e}-Hopf algebras in Majid-Ruegg and classical Poincar\'{e} bases differ 
 only by nonlinear change of the fourmomenta generators.
In both bases the algebraic and coalgebraic sectors are described as $\kappa$-deformed bicrossproduct Hopf algebras.
The choice of bicrossproduct basis appears to be convenient in the study of duality relations between $\kappa$-deformed Poincar\'{e} algebra and  $\kappa$-deformed Poincar\'{e} group (see Sect.~3).

It is known that leading $\frac{1}{\kappa}$ deformation term in the coproducts 
  for quantum-deformed Lie algebras 
  is provided for large class of bases  
by means of  general formula ($\hat{g}$ describes Lie algebra generators)
\begin{equation}\label{jlprag17}
\Delta(\hat{g}) = \Delta^0 (\hat{g}) + \frac{1}{\kappa} \left[
\Delta^{(0)} (\hat{g}), \hat{r}
\right] + O(\frac{1}{\kappa^2})
\end{equation}
where $\hat{r} \in \hat{g} \otimes \hat{g}$ 
denotes the classical $r$-matrix which satisfies the classical Yang-Baxter equation (CYBE)

\begin{equation}\label{jlprag18}
\left[ \left[ \widehat{r}, \widehat{r} \right] \right] 
\equiv
\left[ \widehat{r}_{12}, \widehat{r}_{13} \right]
+
\left[ \widehat{r}_{12}, \widehat{r}_{23} \right]
+
\left[ \widehat{r}_{13}, \widehat{r}_{23} \right]
=
\hat{\Omega}_3 
\equiv
 \Omega^{ijk}_3  I_i \wedge I_j \wedge I_k
\end{equation}
In (\ref{jlprag18}) the expression 
$\left[ \left[ \widehat{r}, \widehat{r} \right] \right] $  describes Schouten bracket
 (see e.g. \cite{jlrprag14}), 
 and
$\Omega_3 \in \hat{g} \otimes \hat{g} \otimes \hat{g}$ is a~$\hat{g}$-invariant 3-form.
One gets the following two classes of deformations:
\\[4pt]
\noindent
a) satisfying standard CYBE with $\Omega_3\equiv 0$.

In such a case it can be shown that the deformation (quantization) of Lie algebra $\hat{g}$ is determined by twist function $\hat{F}\in U(\hat{g})\otimes U(\hat{g})$, determining the deformed coproducts as follows:
\begin{equation}\label{jlprag19bisbis}
\Delta(\hat{g}) = \hat{F}(\hat{g}) \circ \Delta^{(0)}(\hat{g})\circ \hat{F}(\hat{g})
\end{equation}
where $D^0(\hat{g}) = \hat{g} \otimes 1 + 1 \otimes \hat{g}$ and
 $A\circ B = A_{(1)} B_{(1)} \otimes A_{(2)} B_{(2)}$  \, where   $A=A_{(1)}\otimes A_{(2)}$ etc. 
\\[4pt]
\noindent
b) satisfying  modified  CYBE with $\Omega_3 \neq 0$.

It appears that for D=4 Poincar\'{e} algebra we get  unique choice of $\hat{g}$-invariant 3-form,  which may occur in CYBE namely

\begin{equation}\label{jlprag19}
\Omega_3(P_\mu,  M_{\mu\nu}) = t \, P_\mu \wedge P_\nu \wedge M^{\mu\nu}
\qquad
\begin{array}{c}
t \hbox{\, - \, complex parameter}
\end{array}
\end{equation}

The  $\kappa$-deformed Poincar\'{e} algebras obtained in \cite{jlrprag27,jlrprag28} 
are generated by the following classical $r$-matrix (see (\ref{jlprag17}))

\begin{equation}\label{jlprag20}
\hat{r} = N_i \wedge P_i 
\qquad \quad (a\wedge b \equiv a \otimes b - b\otimes a)
\end{equation}
In a middle of  nineties (see e.g. \cite{jlrprag54}) there was introduced the  generalization of   $\kappa$- deformations depending on constant fourvector  $a_\mu$, generated by the following 
 generalization of classical \hbox{$r$-matrix} (\ref{jlprag20})
\begin{equation}\label{jlprag21}
\hat{r}(a_\mu) = \frac{1}{\kappa} P_\mu \wedge M^{\mu\nu} \, a_\nu
\end{equation}
CYBE for the generalized $r$-matrix  (\ref{jlprag21}) takes the following form 

\begin{equation}\label{jlprag22}
\left[ \left[
 \hat{r} (a_\mu ), \hat{r} (a_\nu ) \right] \right] =
a_\mu a^\mu \cdot \Omega_3 (P_\mu, M_{\mu\nu})
\end{equation}
For Lorentzian signature  ($a_\mu a^\mu= \vec{a}^{\, 2} - a^2_0$) one obtains three different
 types of
  $\kappa$-deformations, which can not be related by any change of   Poincar\'{e} algebra basis. 
We get namely that
\\ \\ 
1) If $a_\mu a^\mu =1$ (one can choose $a_\mu = (1,0,0,0)$) the generalized classical $r$-matrix (\ref{jlprag21}) reduces to (\ref{jlprag20}).
We obtain in such a case standard or time-like $\kappa$-deformations (see e.g. (\ref{jlprag11})--(\ref{jlprag16})).
\\[10pt]
2) If $a_\mu a^\mu = -1$ (one can choose $a_\mu= (0,1,0,0)$)   one gets tachyonic $\kappa$-deformation \cite{jlrprag55}.
\\[10pt]
3) If $a_\mu a^\mu =0$ (one can choose light-like vector $a_\mu=(1,1,0,0)$) 
 one obtains the light-cone $\kappa$-deformation \cite{jlrprag56}.
It follows from (\ref{jlprag22}) that for light-cone $\kappa$-deformation the classical $r$-matrix (\ref{jlprag21})
 describes the solution of  standard CYBE, with $\Omega_3 = 0$.
Such quantization  can be realized by twist factor, which has been explicitly
 calculated \cite{jlrprag57,jlrprag58}.

\section{Quantum $\kappa$-Poincar\'{e} group from Hopf-algebraic duality}

If we consider classical symmetries in classical mechanics the finite symmetry transformations  are described by  elements $g$ of classical matrix groups
 $G$ $(g\in G)$, and infinitesimal ones are calculated by using the matrix realizations of corresponding Lie algebra
 $\hat{g}$. 
 The theory of quantum symmetries was analogously  developed in two-fold way.
Firstly,  the classical matrix  groups $G$ after their  consistent  \hbox{$q$-deformation} were used to construct the Hopf algebras $\tilde{\mathds{H}}$ of functions $F(\hat{G}_q)$, 
   with noncommutative elements  of quantum-deformed matrix group  $\hat{G}_q$.
 In second way, Lie algebras $\hat{g}$ and their enveloping algebras $U(\hat{g})$ were deformed into the quantum enveloping algebras $U_q(\hat{g})$.
After supplementing  consistently the Hopf-algebraic operations  (coproducts $\Delta$, antipodes $S$ and counits 
 $\in$)  one gets  quantum Lie-Hopf  algebra ${\mathds{H}}$.
 In previous Section we presented $\mathds{H}$ as $\kappa$-Poincar\'{e}-Hopf algebra describing deformed relativistic space-time symmetries; 
 below we shall introduce quantum $\kappa$-Poincar\'{e} group $\tilde{\mathds{H}}$ as its dual Hopf-algebra.

One says that the Hopf algebra $\mathds{H} = (A=U_q(\hat{g}), m, \Delta, S, \in)$
 is in duality with Hopf  algebra $\tilde{\mathds{H}}= (A^\star = F(\hat{G}_q), m^\star, \Delta^\star, S^\star, \in^\star)$
 if there is a nondegenerate pairing  $< \cdot , \cdot >$: $A\otimes A^\star \to \mathds{C}$ such that 
$(a,b \in A, c, d\in A^\star)$

\begin{subequations}\label{jlprag23}
\begin{eqnarray}
\langle ab, c \rangle = \langle a \otimes b, \Delta^\star (c) \rangle \equiv \langle a, c_{(1)} \rangle \langle b, c_{(2)}\rangle
\label{jlprag23a}
\\[12pt]
\langle \Delta(a), c\otimes d \rangle = \langle a, cd\rangle \equiv \langle a_{(1)}, c \rangle \langle a_{(2)}, d\rangle
\label{jlprag23b}
\end{eqnarray}
\end{subequations}
where $\Delta(a)=a_{(1)} \otimes a_{(2)}$ and $\Delta^\star(c)= c_{(1)} \otimes c_{(2)}$.
Further we postulate that
\begin{eqnarray}\label{jlprag24}
&\langle S(a), c\rangle = \langle a, S^\star (c) \rangle
\nonumber \\[-8pt]
\\\nonumber 
&\langle a, 1_{A^\star} \rangle = \in (a) \qquad  \quad \langle 1_A , c \rangle = \in^\star (c)
\nonumber
\end{eqnarray}

The duality relations (\ref{jlprag23}) link the multiplication (products) in
  $\mathds{H}(\tilde{\mathds{H}})$ and
 comultiplication (coproducts) in $\tilde{\mathds{H}}(\mathds{H})$.
One obtains the following diagram

$$
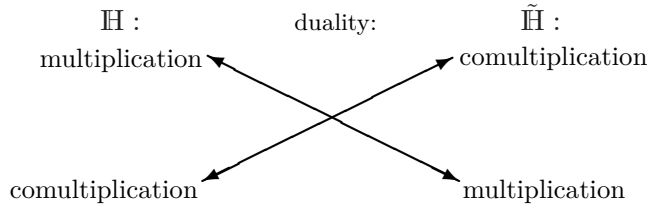
\begin{array}{c}
\phantom{XX} \mathds{H}: \phantom{XXXXXXXXXXXXXX} \tilde{\mathds{H}}: \phantom{XXXXX}
\\[-12pt]
\setlength{\unitlength}{3671sp}%
\begingroup\makeatletter\ifx\SetFigFont\undefined%
\gdef\SetFigFont#1#2#3#4#5{%
  \reset@font\fontsize{#1}{#2pt}%
  \fontfamily{#3}\fontseries{#4}\fontshape{#5}%
  \selectfont}%
\fi\endgroup%
\begin{picture}(4900,1345)(565,-1130)
 \thicklines
{\color[rgb]{0,0,0}\put(3537,-112){\vector( 2, 1){  0}}
\put(3537,-112){\vector(-2,-1){1672}}
}%
\put(2482, 72){\makebox(0,0)[lb]{\smash{{\SetFigFont{9}{10.8}{\rmdefault}{\mddefault}{\updefault}{\color[rgb]{0,0,0}duality:}%
}}}}
\put(781,-179){\makebox(0,0)[lb]{\smash{{\SetFigFont{10}{12.0}{\rmdefault}{\mddefault}{\updefault}{\color[rgb]{0,0,0}multiplication}%
}}}}
\put(3584,-162){\makebox(0,0)[lb]{\smash{{\SetFigFont{10}{12.0}{\rmdefault}{\mddefault}{\updefault}{\color[rgb]{0,0,0}comultiplication}%
}}}}
\put(580,-1062){\makebox(0,0)[lb]{\smash{{\SetFigFont{10}{12.0}{\rmdefault}{\mddefault}{\updefault}{\color[rgb]{0,0,0}comultiplication}%
}}}}
\put(3605,-1057){\makebox(0,0)[lb]{\smash{{\SetFigFont{10}{12.0}{\rmdefault}{\mddefault}{\updefault}{\color[rgb]{0,0,0}multiplication}%
}}}}
{\color[rgb]{0,0,0}\put(1903,-104){\vector(-2, 1){  0}}
\put(1903,-104){\vector( 2,-1){1678.800}}
}%
\end{picture}%
\end{array}
$$

Two Hopf algebras $\mathds{H}$ and $\tilde{\mathds{H}}$ act one on another - we gets the formulae for right action $\triangleright$ and left action $\triangleleft$ (see e.g. \cite{jlrprag14})

\begin{eqnarray}\label{jlprag25}
& a \triangleright c = c_{(1)} \langle a , c_{(2)} \rangle
\nonumber \\[-10pt]
\\\nonumber
& a \triangleleft c = \langle a_{(1)}, c \rangle a_{(2)}
\nonumber
\end{eqnarray}
The action of $A$ on products of elements of  $A^\star$ is given by the Hopf-algebraic formula

\begin{equation}\label{jlprag26}
a \triangleright cd = (a_{(1)} \triangleright c)(a_{(2)} \triangleright d)
\end{equation}
and similarly for the left action on the product in $A$.
 We say that $A^\star$ is a right $\mathds{H}$ module, and $A$ is a left $\tilde{\mathds{H}}$-module.

Using the relations (\ref{jlprag23}b) one can get from the coproducts of standard $\kappa$-deformed
Poincar\'{e}-Hopf algebra the algebra of elements
   $\tilde{e}^B \in A^\star$ ($A=1,\ldots , 10$) 
 spanning the 
  orthonormal dual linear basis, 
 satisfying the relation 
\begin{equation}\label{jlprag27}
\langle e_A, \tilde{e}^{\, B} \rangle = \delta_A^{\, B}
\end{equation}
where $e_A= (P_{\mu}, M_{\mu\nu}) \in A,$ ,  $\tilde{e}^{\, B}= (\hat{x}^\mu, \hat{\Lambda}^{\mu\nu} ) \in \tilde{A}$. One gets explicitly \cite{jlrprag33,jlrprag28,jlrprag29}

\begin{subequations}\label{jlprag28}
\begin{eqnarray}
&& \left[ \hat{x}^\mu, \hat{x}^\nu \right] = - \frac{i}{\kappa} (
\hat{x}^\mu \delta_0^\mu - \hat{x}^\nu \delta^\mu_0
)
\label{jlprag28a}\\\cr
&&\left[ \hat{x}^\mu, \hat{\Lambda}^\nu_\rho \right] = \frac{i}{\kappa}
(
\Lambda^\nu_0 - \delta^\nu_0
) \Lambda^\mu_{\ \rho} 
+ \eta^{\mu\nu} 
(
\hat{\Lambda}^0_{\, \rho} - \delta^0_{\, \rho}
)
\label{jlprag28b}\\\cr 
&&\left[ \hat{\Lambda}^\mu_{\, \nu}, \hat{\Lambda}^\rho_{\, \tau} \right] =0
\label{jlprag28c}
\end{eqnarray}
\end{subequations}

The relation (\ref{jlprag28}a) describes the algebra of  $\kappa$-deformed  D=4 Minkowski space-time  corresponding to the choice $a_\mu= (1,0,0,0)$ of the constant fourvector  introduced in formula~(\ref{jlprag21}).
It should be also observed that due to the rhs of relation (\ref{jlprag28}b) the formulae (\ref{jlprag28}a--c) do not describe Lie algebra with linear commutators.

Similarly, using relations (\ref{jlprag23}a) one obtains the following coproduct  formulae in $\tilde{\mathds{H}}$:
\begin{eqnarray}\label{jlprag29}
&&\Delta(\hat{x}_\mu) = \hat{x}^\mu \otimes 1 + \hat{\Lambda}^\mu_{\ \nu} \otimes \hat{x}^\nu
\nonumber\\ \\[-10pt]
&&\Delta(\hat{\Lambda}_{\mu}^{\ \nu}) = \hat{\Lambda}_\mu^{\ \rho} \otimes \hat{\Lambda}_\rho^{\ \nu}
\nonumber
\end{eqnarray}
The coproducts  (\ref{jlprag29}) describe the undeformed Poincar\'{e} group transformations.
 Indeed, if we put $\hat{b}^\mu=\hat{x}_\mu \otimes 1$, $\hat{\alpha}^\mu_{\ \nu} =\hat{\Lambda}_\mu^{\ \nu} \otimes 1$ and $\Delta(\hat{x}_\mu) = \hat{x}'_\mu$, $1\otimes \hat{x}_\mu=\hat{x}_\mu$ one gets the  standard  transformation laws
\begin{eqnarray}\label{jlprag30}
& \hat{x}'_\mu = \hat{b}_\mu + \hat{\alpha}_\mu^{\ \nu} \hat{x}_\nu
\nonumber \\ \\[-10pt]
&  \Lambda'^{\ \nu}_\mu = \hat{\alpha}_{\mu}^{\ \rho} \hat{\Lambda}_{\rho}^{\ \nu}
\nonumber
\end{eqnarray}
It should be observed that following \cite{jlrprag13,jlrprag16} the classical transformation laws (coproducts) remain  valid after the quantum 
 deformation (quantization) $G \to \hat{G}_q$ of  any classical matrix group  $G$. 
 Denoting  noncommutative matrix entries $\hat{G}^a_{\ b} \in \hat{G}_q$, we get general coproduct formula 
 for quantum matrix groups 
\begin{equation}\label{jlprag31}
\Delta(\hat{G}^a_{\ b}) = \hat{G}^a_{\ c} \otimes \hat{G}^c_{\ b}
\end{equation}

The pair of dual Hopf algebras can be calculated as well for generalized $\kappa$-deformations, described by the classical $r$-matrix (\ref{jlprag21}).
In particular one gets the following $\kappa$-deformed Minkowski space  algebra depending on constant  fourvector $a_\mu$   
 \begin{equation}\label{jlprag32}
\begin{array}{c}
\hbox{generalized}
\\
\hbox{$\kappa$-deformed}
\\
\hbox{Minkowski space:}
\end{array}
\qquad 
\left[ \hat{x}^\mu , \hat{x}^\nu  \right] = \frac{i}{\kappa}
(\hat{x}^\mu  a^\nu - \hat{x}^\nu  a^\mu)
\end{equation}
It appears that the noncommutativity in quantum space-time (\ref{jlprag32}) is present in only one quantum coordinate 
$\hat{x}^\mu a_\mu$; if we introduce three linearly independent fourvectors 
$b^{(i)}_\mu$ ($i=1,2,3$) which are orthogonal to
   $a^\mu $  (i.e. $a^\mu b^{(i)}_\mu= 0$),  
	 the remaining three quantum coordinates $\hat{x}^{(i)}=b^{(i)}_\mu \hat{x}^{\mu}$ describe three-dimensional commutative   manifold, 
	 i.e. $[ \hat{x}^{(i)}, \hat{x}^{(j)}] =0$ for any $i,j$.
	
	The dual pair of Hopf algebras $\mathds{H}, \tilde{\mathds{H}}$ is useful in describing the algebraic quantum-deformed phase spaces.
	For that purpose one introduces cross-products of two dual Hopf algebras (see e.g. \cite{jlrprag14})
	\begin{equation}\label{jlprag33}
	{\mathcal{H}} = \mathds{H} \ltimes \tilde{\mathds{H}} \qquad \qquad\qquad
	({\rm{e.g.}} \ U_\kappa(\hat{g}) \in \mathds{H}, \  F(\hat{G}_\kappa) \in \tilde{\mathds{H}})
	\end{equation}
	which is called Heisenberg double algebra $\mathcal{H}$.
	The cross product multiplication rule for $a\in \mathds{H}$ and $c\in \tilde{\mathds{H}}$ is given by formula
	($\mathcal{H} = \mathcal{A} \otimes \mathcal{A}^\star$)
	\begin{equation}\label{jlprag34}
	a \cdot c \equiv (a \otimes 1) \cdot (1 \otimes c) = c_{(1)} \langle a_{(1)}, c_{(2)} \rangle a_{(2)}
	\end{equation}
	what permits to calculate in $\mathcal{H}$ as well the cross commutators between $\mathds{H}$ 
	 and $\tilde{\mathds{H}}$.
	A simple  application of  Heisenberg  double   construction  is provided by  relativistic quantum-mechanical Heisenberg algebra.
	Such simple Heisenberg double is obtained from  two dual Abelian Hopf algebras describing commuting space-time coordinates $\hat{x}_\mu \in \tilde{\mathds{H}}_x$ and fourmomenta $\hat{p}_\mu \in {\mathds{H}}_p$; the Heisenberg double
	$\mathcal{H}^{(0)}_{4,4} = {\mathds{H}}_p \ltimes \tilde{\mathds{H}}_x$  is characterized by  the cross-product  relations derived from (\ref{jlprag34})
	 (we put $\hbar=1$) 
	\begin{equation}\label{jlprag35}
	\left[ \hat{x}_\mu , \hat{p}^\nu \right] = i \, \delta_\mu^{\ \nu}
	\end{equation}
	i.e. we get    relativistic quantum-mechanical phase space. 
	Subsequently, various quantum-deformed phase spaces can be treated as consistent deformations of  the classical Heisenberg double $\mathcal{H}^{(0)}_{4,4}$.
	
	The Heisenberg  doubles describe quantum spaces, however without Hopf-algebraic structure.
	In Sect.~5  we shall show that Heisenberg doubles $\mathcal{H}$ define $\mathds{H}$-covariant quantum-deformed phase spaces, which 
	  belong to the algebraic category of  quantum spaces with	 Hopf algebroid structure.

	\section{From  $\kappa$-deformed Hopf-algebraic symmetries to DSR approach and curved 
	momentum space}
	
	In search for  an explanation of possible high energy effects of QG (see e.g. \cite{jlrprag59})
	 there was proposed around year 2000 the generalization of  Einstein special relativity, with two observer-independent parameters - light velocity $c$ and Planck length $L_p \simeq {\rm 10}^{-33}$cm - named  Doubly Special  Relativity (DSR) \cite{jlrprag60}--\cite{jlrprag63}\footnote{Later the notation ``DSR'' was used as   well in more general sense, with the meaning ``Deformed Special Relativity''.}.
	
	In first formulation of DSR framework \cite{jlrprag60,jlrprag63} the basic formulae were coincident  with the ones which follow from the formulation of $\kappa$-deformed Poincar\'{e} algebra in bicrossproduct basis, with supplementary identification of mass-like  deformation parameter $\kappa$ as the Planck mass $m$\footnote{For $\hbar=c=1$ one gets $m_p= (L_p)^{-1}$.}.
	\\[6pt]
	The basic two notions of DSR framework are the following
	\\[6pt]
	i) Deformed energy-momentum dispersion relation, which was called in \cite{jlrprag60} ``key characteristic of DSR, both conceptually and phenomenologically''.
	The deformed mass shell condition presented in \cite{jlrprag60} was earlier described  by the formula for $\kappa$-deformed mass Casimir in bicrossproduct basis \cite{jlrprag28,jlrprag29}.
	Further it can be shown (see e.g. \cite{jlrprag64}) that various $\kappa$-deformed mass-shell conditions in DSR framework were directly  related with $\kappa$-deformed Poincar\'{e}-Hopf algebra written in different algebra bases.
	\\[6pt]
	ii) Deformation of  classical linear Lorentzian boost transformations of fourmomenta into the
	 ones described by nonlinear formulae \cite{jlrprag65}.
	We add that DSR framework, by following the description of particle kinematics in scattering theory,  is  formulated in commutative fourmomentum space,  with   commuting space-time coordinates  introduced by standard Fourier transform of the momentum-dependent functions (see e.g. \cite{jlrprag66})\footnote{Analogous
	naive way of introducing space-time coordinates  was used in early formalism of $\kappa$-deformed  relativistic symmetries, in 1991--1993 \cite{jlrprag26,jlrprag27,jlrprag29,jlrprag67}. 
	Firstly it was clearly stressed in \cite{jlrprag28} that the modules of $\kappa$-deformed Poincar\'{e}-Hopf algebra, e.g. describing $\kappa$-Minkowski spaces, should be necessarily noncommutative.}.
	It was shown however \cite{jlrprag68,jlrprag69} that the modified finite Lorentz transformations calculated in \cite{jlrprag65} can be also obtained by 
	 using the 
	 Hopf algebra structure of  $\kappa$-deformed Poincar\'{e}  algebra in bicrossproduct basis.
	The general Hopf-algebraic expression for finite $\kappa$-deformed boosts in arbitrary basis is given by the  following  formula (see \cite{jlrprag70}, Sect.~2d)\footnote{We choose boosts along the third space axis.}.
	
	\begin{eqnarray}\label{jlprag38}
	P_\mu (\alpha) =  ad _{e^{i\alpha N_3}} P_\mu \equiv \sum\limits^{\infty}_{k=0}
	  \frac{i^k}{k!}
	 \ (ad_{\alpha N_3} (ad_{\alpha N_3} \ldots (ad_{\alpha N_3} P_\mu ) \ldots ))
	\end{eqnarray}
	where quantum adjoint action $ad_{Y} {X}$ defined as follows
	\begin{equation}\label{jlprag39}
	ad_{Y} X = Y_{(1)} XS(Y_{(2)})  \qquad \qquad (\Delta(Y) = Y_{(1)} \otimes Y_{(2)})
	\end{equation}
	can be expressed in Majid-Ruegg bicrossproduct basis  of 
	 $\mathcal{U}_\kappa(\hat{p}_{3,1})$ by known ``classical'' formula
	
	\begin{equation}\label{jlprag40}
	P_\mu(\alpha) = e^{i\alpha N_3}\, P_\mu \, e^{-i\alpha N_3}
	\end{equation}
	The differential equation following from formula (\ref{jlprag40}) was used in \cite{jlrprag65} in order to calculate explicitly  the nonlinear boost transformations  in DSR theory.
	
	 Important question addressed as well by DSR approach is $\kappa$-deformed addition of fourmomenta describing modified conservation law.
	In Hopf-algebraic approach such addition is determined by the coproducts and 
	 from quantum algebra/quantum group
	duality
	 follows the appearance  of noncommutative quantum space-times.
	In DSR approach
	   the choice of coproduct was rather ambiguous, because the  symmetric and nonsymmetric coproducts, both allowed by deformed mass Casimirs, were used.
	In the  case of symmetric addition law (see e.g. \cite{jlrprag63}, where  the postulates of DSR theory are 
	discussed) the deformed relativistic symmetries 
	 of fourmomenta 
	 are described by the standard special relativity  rules,  however  expressed by   nonlinearly  transformed  classical fourmomenta generators   (see \cite{jlrprag68}, 
	\cite{jlrprag71}--\cite{jlrprag73}).
	If we employ however the deformed  Poincar\'{e} algebra as quantum group\footnote{See Drinfeld' definition of quantum group in \cite{jlrprag12}, described as noncocommutative Hopf algebras with nonsymmetric coproducts ($\Delta(x) = \Delta_{(1)} (x) \otimes \Delta_{(2)}(x) \neq \Delta_{(2)}(x) \otimes \Delta_{(1)}(x)$).} the coproducts are necessarily nonsymmetric and dual space-time becomes noncommutative.
	It appears that for  some large class of  $\kappa$-Poincar\'{e} algebra bases the  \hbox{$\kappa$-deformed} quantum space-time is described by $\kappa$-Minkowski space  (see also (\ref{jlprag30}a)).
	
	In DSR formalism instead of full Hopf-algebraic description of $\kappa$-deformed relativistic symmetries one restricts usually the framework to the so-called DSR algebras \cite{jlrprag74}--\cite{jlrprag77} describing
	 \hbox{$\kappa$-deformation} of classical semidirect product $R_{3,1} \rtimes \hat{p}_{3;1}$, where $R_{3,1}$ denotes Minkowski space and
	$\hat{p}_{3;1}$ the classical Poincar\'{e} algebra\footnote{Such algebras were also called Heisenberg-Poincar\'{e} algebras \cite{jlrprag78}.}.
	 DSR algebra is spanned by generators ($\hat{X}_\mu, \hat{P}_\mu, \hat{M}_{\mu\nu}$), 
	with $\hat{X}_\mu$ describing $\kappa$-deformed Minkowski space and 
	$(\hat{P}_\mu , \hat{M}_{\mu\nu})$ given by quantum-deformed Poincar\'{e} algebra which can be
	 endowed with Hopf algebra structure.
	In such algebraic (not Hopf-algebraic!)  approach the deformations of DSR algebra are present  in the commutators 
	${ [ \hat{X}_\mu, \hat{P}_\nu ] }$ and $[ \hat{X}_\mu, \hat{M}_{\rho, \tau}]$
	 ( see (\ref{jlprag47})).                      
	
	Because fourmomenta $P_\mu$ are Abelian, one can look also for the realizations of DSR algebra in terms of classical Heisenberg algebra generators $\hat{p}_\mu , \hat{x}_\mu$, satisfying the relations (\ref{jlprag35}).
	If we use the Schr\"{o}dinger realization of the algebra (\ref{jlprag35})
	\begin{equation}\label{jlprag41}
	\hat{p}_\mu = p_\mu \qquad \qquad \hat{x}_\nu = x_\nu \equiv i \frac{\partial}{\partial p^\nu}
	\end{equation}
	one can  express the $\kappa$-Minkowski coordinates $\hat{X}_\mu $  by using  of momentum-dependent tetrad
	 $E^\alpha_\mu(p)$
	\begin{equation}\label{jlprag42b}
	\hat{X}_\mu = \hat{x}_\alpha \, E^\alpha_\mu (p)
	\end{equation}
	with $E^\alpha_\mu (p)$ determined by the commutator $[ \hat{X}_\mu , \hat{P}_\mu ]$.
	Embedding of $\kappa$-deformed quantum symmetry algebras
	 with Hopf algebra structure 
	 into enveloping classical Heisenberg algebra has been also  used, however  such procedure is questionable  
	 because such a realization 
	  does not preserve the Hopf-algebraic structure\footnote{The inconsistency follows from the fact that Heisenberg algebra (\ref{jlprag35}) is not endowed with bialgebra structure (see Sect.~5.1). An example of   realization of  Hopf algebra as embedding  into enveloping Heisenberg algebra can be found e.g. in
	\cite{jlrprag79}.}.
	
	The non-Abelian  fourmomenta coproducts, if applied to the description of   two  fourvectors $p_\mu$, 
	and $dp_\mu$ indicate the curved structure of fourmomentum space (see e.g. \cite{jlrprag80,jlrprag81}). 
	In DSR approach instead of using coproducts and algebraic methods of  Hopf algebra theory, the curved momentum space  techniques are used as representing the non-Abelian fourmomenta addition law.
	It was realized that coproducts of  Poincar\'{e}-Hopf algebras with associative composition law of fourmomenta  describe curved Cartan-Riemann fourmomentum spaces, with nonvanishing torsion and vanishing curvature.
	The examples of such curved spaces are provided by group manifolds.
	Indeed, for $\kappa$-deformed Poincar\'{e}-Hopf algebra it was shown \cite{jlrprag82}--\cite{jlrprag84} that corresponding curved fourmomentum space is described by a four-dimensional Lie group $A\cdot N(3)$, entering into the Iwasawa decomposition of $SO(1,4)$ group manifold.
	
	The use of nonlinear momentum space with nontrivial metric which leads to non-Abelian composition   law  leads to nonstandard 
	description of dynamical systems. 
	Such framework helped to introduce the notion of relative locality \cite{jlrprag85} providing space-time as derived concept which is described effectively by the interactions of particle probes in momentum space. 
	 It can be also added that recently the curved momentum space has been incorporated into dynamical
	 curved phase space framework describing so-called meta-string theory 
		 (see  e.g. \cite{jlrprag86,jlrprag87}).

\section{Recent applications of  $\kappa$-deformations - two examples }

The applications of $\kappa$-deformed symmetries to physical models (e.g.~\!deformed QFT)  are somewhat limited because for $\kappa$-deformed  Poincar\'{e}-Hopf algebra the universal $R$-matrix is not known.
We shall provide two recent results in the framework of  $\kappa$-deformations which use the $\kappa$-deformed Heisenberg double (see \cite{jlrprag88}) and classical $\kappa$-Poincar\'{e} $r$-matrices
 (see formulae (\ref{jlprag20}),(\ref{jlprag21})).

\subsection{$\kappa$-deformed phase space as Hopf bialgebroid}

It has been recently  observed \cite{jlrprag89}--\cite{jlrprag92}, \cite{jlrprag44,jlrprag45} that quantum phase spaces can be supplemented  with coalgebra structure in the framework of Hopf algebroids \cite{jlrprag93,jlrprag94}.
It is known already a long-time that large class of quantum-deformed phase spaces, with
coordinate sector  described by Hopf-algebraic quantum group, can be identified as Heisenberg double  \cite{jlrprag88}.
Further, exploiting Liu theorem\footnote{This theorem is proved rigorously for finite-dimensional Hopf algebras, but formally it can be applied to infinite-dimensional case \cite{jlrprag45,jlrprag92}.} that finite-dimensional Heisenberg double has the structure of a Hopf  
 algebroid~\cite{jlrprag94}, there was shown \cite{jlrprag44,jlrprag45} that quantum $\kappa$-deformed phase space is the quantum space with Hopf algebroid structure.

While Hopf algebras are quantum analogues of groups, Hopf algebroids are quantum analogues of  grupoids \cite{jlrprag95}.
Hopf algebroids are bialgebroids supplemented with the antipode map and additional structures in algebra sector:
base subalgebra $B$ of total algebra $A$ and two maps ($h\in A, b\in B$)
\\[4pt]
i) Source algebra map $s$: \  $b\cdot h = s(b)h $  \  $s(b)\in h$
\\[4pt]
ii) target antialgebra map $t$: $h\cdot b = t(b)h$ \ $t(b)\in h$
\\[4pt]
which commute, i.e.
\begin{equation}\label{jlprag43}
[ s(b) , t(b') ] = 0 \qquad \qquad b,b' \in B
\end{equation}
The Hopf algebroid $\mathds{H}$ is specified as follows:
\begin{equation}\label{jlprag44}
{\mathds{H}} = \left(
A, m; B, s, t; \Delta, s, \in
\right)
\end{equation}
where $B$ in the case of Hopf algebra   reduces to the unity element $\mathds{1}$ of algebra $A$, i.e. disappears.

The role of base algebra plays essential role in defining coalgebraic sector of bialgebroid: one defines coproducts $\Delta$ using new tensor product $A \mathop{\otimes}\limits_{\tiny{\hbox{B}}}A$ over generally noncommutative base  ring  $B$ \cite{jlrprag96,jlrprag97}.
The coproduct for bialgebroid can be also realized in terms of standard tensor  products $A \otimes A$.
If we introduce left ideal  in algebra $A$ defined in terms of source and target maps ($s,t$) as follows
\begin{equation}\label{jlprag45}
I_L = {s}(b) \otimes 1 - 1 \otimes {t}(b)
\end{equation}
the tensor product  $A \mathop{\otimes}\limits_{\tiny{\hbox{B}}}A$ is defined by the following equivalence classes of standard tensor products  $A\otimes A$  (see e.g. \cite{jlrprag97})
\begin{equation}\label{jlprag46}
h \mathop{\otimes}\limits_{\small{\hbox{B}}}  h' \simeq
  h 
\otimes  h'   \  \ 
  \hbox{iff} \ \  I_L \circ (h\otimes h')  = s(b)h \otimes h' - h \otimes t(b) h'  =0
\end{equation}
for all $b\in B$. The ideal $I_L$ generates
 in terms of standard tensor product
 the nonuniqueness, which is consistent with the homomorphism property of coalgebra  reproducing  the algebraic structure of $A$.
In \cite{jlrprag44}  such a freedom was called the coproduct gauge.

The simplest example of Hopf bialgebroid is  a standard quantum phase space with canonical Heisenberg  relations 
(see  (\ref{jlprag35})), which is given as a Heisenberg double of a pair of Abelian and co-Abelian Hopf algebras $\tilde{\mathds{H}}_x$ and ${{\mathds{H}}}_p$, describing classical coordinates $x_\mu$ and classical momenta 
$p^\mu$.
The cross-product multiplication rule (\ref{jlprag34}) applied to Heisenberg double $\mathds{H}_p \ltimes \mathds{H}_x$ leads to the relation (\ref{jlprag35}), i.e. results in the quantization of Poisson structure on classical phase space.
If we calculate the Heisenberg double of $\kappa$-deformed Poincar\'{e}-Hopf algebra $\mathds{H}_\kappa$  and dual quantum 
 \hbox{$\kappa$-Poincar\'{e}} group $\tilde{\mathds{H}}_\kappa$, with generators ($P_\mu, M_{\mu\nu}, \hat{x}_\mu , \hat{\Lambda}_{\mu\nu}$), we obtain the following cross-commutators  (see \cite{jlrprag88})
\\[4pt]
i) $\kappa$-deformed relations containing $\hat{x}_\mu$
\begin{eqnarray}\label{jlprag47}
&&\left[ P_\mu , \widehat{x}_\rho \right] = - i \, \eta_{\mu\rho} - \frac{i}{\kappa} 
		(\eta_{\rho 0} P_\mu - \eta_{\mu 0} P_\rho)
		\nonumber\\ \\
		&& \left[ M_{\mu\nu}, \widehat{x}_\rho \right] = i (\eta_{\mu\rho} \widehat{x}_\nu - \eta_{\nu\rho} \widehat{x}_\mu) 
		+ \frac{i}{\kappa} (\eta_{0\mu} M_{\nu\rho} - \eta_{0\nu} M_{\mu\rho})
\nonumber
\end{eqnarray}
ii) nondeformed relations containing

\begin{eqnarray}\label{jlprag48}
&&  [  P_\mu , \hat{\Lambda}_{\nu\rho} ] = - i (\eta_{\mu\nu} P_\rho
- \eta_{\nu\rho} P_\nu )
\nonumber\\ \\
&& [ M_{\mu\nu}, \hat{\Lambda}_{\rho\tau} ] = i (\eta_{\mu\rho} \hat{\Lambda}_{\nu\tau} 
- \eta_{\nu\rho} \hat{\Lambda}_{\mu\tau}) - (\rho\leftrightarrow \tau)
\nonumber
\end{eqnarray}
The relations (\ref{jlprag47},\ref{jlprag48})  supplemented with the algebraic relations of $\kappa$-deformed Poincar\'{e} algebra $\hat{\mathds{H}}_\kappa$
 (see (\ref{jlprag11})) and $\kappa$-deformed Poincar\'{e} group $\mathds{H}_\kappa$ (see (\ref{jlprag28})) describe generalized $\kappa$-deformed quantum phase space $\mathcal{P}^{10;10}$.
It can be shown that  $\mathcal{P}^{10;10}$ is a Hopf algebra $\hat{\mathds{H}}_\kappa$-module, or equivalently $\mathcal{P}^{10;10}$. is $\kappa$-covariant.

One can distinguish the following three subalgebras of  $\mathcal{P}^{10;10}$.
\\[4pt]
i) $\kappa$-deformed DSR algebra $\mathcal{P}^{10;4}$,  with base ($P_\mu, M_{\mu\nu}; \hat{x}_\nu$), 
describing noncommutative \hbox{$\kappa$-Minkowski} space with acting covariantly the \hbox{$\kappa$-deformed}
 Poincar\'{e}-Hopf algebra ${\mathds{H}}_\kappa$
\\[4pt]
ii) $\kappa$-deformed DSR group $\mathcal{P}^{4;10}$, with base ($\hat{P}_\mu ; \hat{x}^\mu, \hat{\Lambda}^\mu_{\ \nu}$), which is the quantum $\kappa$-Poincar\'{e} group $\mathds{H}_\kappa$ with  
  $\hat{p}_\mu$ acting in covariant way
\\[4pt]
iii) $\kappa$-deformed standard phase space $\mathcal{P}^{4;4}$, with base $(\hat{x}_\mu , \hat{p}_\nu)$, where 
 $\hat{x}_\mu$ satisfy the algebra (\ref{jlprag30}a)  and commutative fourmomenta $\hat{p}_\mu$ have the following cross relations: $(\mu =0, i ; i=1,2,3)$
\begin{eqnarray}\label{jlprag49}
&& [ \hat{p}_i , \hat{x}_j ] = -  i\, \delta_{ij} \qquad \qquad
 [ \hat{p}_i , \hat{x}_0 ] = 0
\nonumber \\  \\
&&[ \hat{p}_0 , \hat{x}_0 ]= -i
\qquad \qquad
 [ \hat{p}_0 , \hat{x}_i ] = \frac{i}{\kappa} \hat{p}_i
\nonumber
\end{eqnarray}
We see that only last relation in (\ref{jlprag49}) is $\kappa$-deformed, in comparison with  standard relativistic Heisenberg algebra (\ref{jlprag35}).

The  generalized phase space $\mathcal{P}^{10;10}$ (with Lorentz spin sector 
($M_{\mu\nu}, \hat{\Lambda}_{\mu}^{\ \nu}$)) as well as  
 $\mathcal{P}^{4;4}$  are endowed with Hopf  algebroid structure.
The base algebra $B$ is identified with quantum group algebra part ($\hat{x}_\mu$ for  $\mathcal{P}^{4;4}$  and ($\hat{x}_\mu , \hat{\Lambda}_\mu^{\ \nu}$) for  $\mathcal{P}^{10;10}$).

The explicite Hopf algebroid structure for standard $\kappa$-deformed quantum phase space  $\mathcal{P}^{4;4}$
 has been obtained by purely algebraic considerations in \cite{jlrprag44,jlrprag45}.
Following 
  Lu theorem  (see \cite{jlrprag94}) we have chosen the  primary coproduct on base algebra as $\Delta(\hat{x}_\mu) = 1 \otimes \hat{x}_\mu$,
	with nonuniqueness described by coproduct gauge, e.g.
\begin{equation}\label{ljprag50}
\Delta(\hat{x}^\mu)
\begin{array}{c}
      \hbox{\small{coproduct}}
      \\[-1pt]
      {\overrightarrow{\ \
      {\small\hbox{gauge}}
      \ \ }}
      \end{array}
			\Delta (\hat{x}{}^\mu ) + \Lambda ^\mu (\hat{x}, \hat{p})
		= \hat{x}{}_\mu \otimes f_\mu^{\ \nu} ({\hat{\vec{p}}})
      \end{equation}
where $f_\mu^{\ \nu}(\hat{p})$ can be calculated by using for coproducts the algebraic structure (\ref{jlprag49}).
 One gets
\begin{eqnarray}\label{jlprag51}
&&\Lambda_i(\hat{x}, \hat{p}) = \hat{x}_i \otimes e^{\frac{P_0}{\kappa}}
			 - 1 \otimes \hat{x}_i
			\nonumber\\ \\
			&&\Lambda_0(\hat{x}, \hat{p}) = \hat{x}_0 \otimes
			1 + \frac{1}{\kappa} \hat{x}_i \otimes
			e^{\frac{P_0}{\kappa}} \hat{x}_i 
			 - 1 \otimes \hat{x}_0
			\nonumber
\end{eqnarray}
The coproduct gauge (\ref{jlprag51}) can be further generalized, in accordance with the  homomorphism 
 of coproducts with the algebraic relations  in $\mathcal{P}^{4;4}$  (see \cite{jlrprag44}).

I would like to add that explicite Hopf algebroid structure of $\mathcal{P}^{10;10}$, with the description of coproduct gauges, is now under considerations.

\subsection{$\kappa$-deformation of string target (super)spaces}

The dynamics of  (super) strings can be described by twodimensional $\sigma$-models, on suitable (super)group or 
(super)coset manifolds.

In particular the action describing D=10 Green-Schwarz superstring  has been  described as D=2 $\sigma$-model with D=10 N=2 super-Poincar\'{e} target manifold \cite{jlrprag98}.
If one replaces in target superspace the D=10 Poincar\'{e} manifold by $AdS_5\times S_5$, with $S_5$ describing an internal  sector, one should consider D=2 $\sigma$-model on supercoset $\mathds{K}$ \cite{jlrprag99}.

\begin{equation}\label{jlprag52}
\begin{array}{c}
AdS_5 \times S_5 \quad  {\rm D=10} \quad \hbox{superstring}
\end{array}
\qquad 
\begin{array}{c}
\longleftrightarrow 
\end{array}
\quad
\begin{array}{c}
 \hbox{D=2 \ $\sigma$-model}
\\
\hbox{supercoset} \ {\mathds{K}}= \frac{PSU(2,2;4)}{O(4,1)\times O(5)}
\end{array}
\end{equation}
The $\sigma$-model action for  strings with target space described by group manifold $G$ or coset $\frac{G}{H}$
 can be expressed by bilinars  in D=2 Lie algebra - valued currents ${\mathds{J}}_\alpha = J^a_\alpha I^a$
 ($\alpha = 1,2$),  where $I^a$ describe the Lie algebra generators
\begin{equation}\label{jlprag53}
S = - \frac{T}{2} \int d^2 \xi \ (h^{\alpha\beta} - \in^{\alpha\beta}) Tr [ {\mathds{J}}_\alpha P {\mathds{J}}_\beta ]
\end{equation}
The trace is taken over matrix values of  the generators $I^a$ and $P$ describes the projection operator $G \to \frac{G}{H}$.
In the presence of quantum deformations described by classical $r$-matrix $\hat{r}= \hat{a}_i \wedge \hat{b}_i
 \in  \hat{g} \otimes \hat{g} $  one can introduce the deformation of string action (\ref{jlprag53}) by introducing Yang-Baxter YB $\sigma$-model \cite{jlrprag100,jlrprag101} with particular choice of  deforming YB kernel $(1-\xi R)^{-1}$
\begin{equation}\label{jlprag54}
S_{v_B} = -\frac{T}{2} \int d^2 \xi (h^{\alpha \beta} - 
\in^{\alpha\beta}) Tr(\mathds{J}_\alpha 
\frac{1}{1-\xi {{R}}}
 P \mathds{J}_\beta)
\end{equation}
The operator $\mathds{R} = R(I_k)\cdot I_k$  is expressed by classical $r$-matrix as follows

\begin{equation}\label{ljprag55}
R(I_k) = \sum \hat{a}_i \langle b_i , I_k\rangle =Str (\hat{r} \circ (1\otimes I_k) )
\end{equation}
and satisfies modified CYBE in Semenov-Tien-Shansky form \cite{jlrprag102,jlrprag103}.
\begin{equation}\label{jlprag56}
[ R(I_j), R(I_k) ]  - {R} [R(I_j), I_k ] + [ I_j, R(I_k) ] = c [I_j, I_k ]
\end{equation}
We add that if we consider   superstrings, the generators ${\mathds {J}}_\alpha $ describe both sets of bosonic
 (even) and fermionic (odd) currents, i.e. ${\mathds{J}}_\alpha \to (J^B_\alpha , J^F_\alpha)$,  and the trace $Tr$ should be replaced in formulae 
(\ref{jlprag53}--\ref{jlprag54}) by supertrace Str. Further if we consider  GS superstring with supercoset (\ref{jlprag52}) as target     space, one should introduce four projection operators, corresponding to $Z_4$-graded decomposition  of  $PSU(2,2;4)$ supergroup generators \cite{jlrprag104}--\cite{jlrprag106}.

Because the conformal algebra $SU(2,2)$ contains as subalgebras the commuting fourmomenta as well as D=4 
Poincar\'{e} algebra, one can embedd D=4 Minkowski space-time into the coset $\mathds{K}$ (see (\ref{jlprag52}).
One can ask further how the target superspace $\mathds{K}$ will be modified and which NS-NS and R-R fluxes will be obtained if we insert in YB $\sigma$-model the classical $r$-matrices (\ref{jlprag20})  or (\ref{jlprag21}).
 Using suitable computer program  such calculation  have been  performed (\cite{jlrprag107}; 
  see also \cite{jlrprag108}).
We calculated YB Minkowski space deformations for standard $\kappa$-deformation (see (\ref{jlprag21}), with $a^2 > 0$),  for tachyonic deformation ($a^2 < 0$)  and for light-cone deformation ($a^2=0$).
We obtained the following results (modulo Abelian $T$-duality transformations \cite{jlrprag109,jlrprag110}):

\begin{table}[h]
\centering
\begin{tabular}{l|l|c}
\hline
type of $\kappa$-def.
& space-time geometry
& remarks
\\
\hline
\\[-8pt]
$a^2> 0 $ (standard)
&D=4 de-Sitter
& $\kappa^{-1}$ describes
\\
&& dS radius
\\
\hline
\\[-8pt]
$a^2 < 0$ (tachyonic)
& D=4 AdS
& $\kappa^{-1}$ describes
\\
&& AdS radius
\\
\hline
\\[-10pt]
$a^2 = 0$ (light-cone)
& $p-p$ wave background
& $x_{+}$ - dependent  \tablefootnote{$x_+ = x_0 + x_3$ describes light-cone time.}
\\
&& {nonvanishing Ricci}
\\
&& curvature; $R=0$
\end{tabular}
\end{table}

It can be added that only in light-cone case one gets nonvanishing  antisymmetric tensor field $B$  (NS-NS 2-form), however in the form of  a total derivative.

An interesting issue is the classification of integrable deformations of $AdS_5 \times S^5$ superstring.
In \cite{jlrprag105} it has been shown that the generalized $a_\mu$-dependent $\kappa$-deformations are integrable.
For three types of $\kappa$-deformations we obtained the explicite formulae for Lax pairs, generating  via Lax equations the infinite number of conserved currents.
In particular it has been shown that zero curvature conditions  for Lax pairs leads to the equations of motion for 
 $\kappa$-deformed YB $\sigma$-model.

An important application of YB $\sigma$-models is the generalization of standard IIB SUGRA theory \cite{jlrprag111}.
It appears that $\kappa$-invariance of D=10 IIB strings embrace as well some more  extensions  of  IIB SUGRA solutions, which can be obtained from YB $\sigma$-models \cite{jlrprag112}.

\section{Final remarks}
I would like to stress that there were presented only some aspects of the theoretical developments related with $\kappa$-deformations of relativistic space-time symmetries and we regret that the list of references is far from being complete.
Let  me  mention  some other important developments:
\\[4pt]
i) SUSY extension of $\kappa$-deformed framework.

It was shown already in 1993 \cite{jlrprag113,jlrprag114} that $\kappa$-deformation of N=1 Poincar\'{e} superalgebra can be calculated by considering quantum WI contraction of  $q$-deformed $OSp(4|1)$ superalgebra.
\\[6pt]
ii) $\kappa$-deformed QM.
\\[4pt]
It has been realized that quantum-deformed particle model in NC phase space $(\hat{x}_\mu, \hat{p}_\mu)$ 
 with  classical momenta sector $(\hat{p}_\mu = p_\mu)$ 
is equivalent via noncanonical map $\hat{x}_\mu = \hat{F}_\mu (x,p)$ 
 to  the free particle model in standard quantum phase space $(x_\mu, p_\mu)$ (for $\kappa$-deformed model see e.g. \cite{jlrprag115}).
For standard time-like $\kappa$-deformations such noncanonical map is well-known from early nineties ($\hat{x}_i = x_i, \hat{x}_0 = x_0 - \frac{1}{\kappa} \vec{p}{\, \vec{x}}$).
\\[6pt]
iii) $\kappa$-deformed classical and quantum field theory.
\\[4pt]
If  one introduces $\star$-product describing the algebra of NC functions on $\kappa$-Minkowski space  
  in terms of classical fields
 \cite{jlrprag116} one can only obtain in such a way the $\kappa$-deformed classical theory.
In quantum $\kappa$-deformed QFT  the algebra of field oscillators is  however $\kappa$-deformed as well 
 \cite{jlrprag117}--\cite{jlrprag119} i.e. in construction of $\kappa$-deformed QFT one should introduce 
 NCQFT with
 new notion of $\star$-product which takes into consideration as well the $\kappa$-deformation of oscillators algebra (see e.g. \cite{jlrprag120}).
\\[6pt]
iv) $\kappa$-deformed gravity action. 
\\[4pt]
Seiberg-Witten map \cite{jlrprag121} applied to $\kappa$-deformed Einstein gravity, with local diffeomorphisms as local
 gauge group, produced higher order curvature corrections \cite{jlrprag122,jlrprag123}.
The lowest nonvanishing  order in $\frac{1}{\kappa}$ is second, i.e. one gets terms  proportional to $\frac{1}{\kappa^2}$.
If we put $\kappa=m_p$, these second order corrections are beyond observability limits in present experiments.

At present still the final choice of QG model is not established.
 Breaking through will arrive  if we shall find experimentally some observable QG effects,  what however still did not happened.
 At present we are therefore  rather at  early  stage of the construction procedure of theoretically sound and phenomenologically confirmed  quantum gravity model.

\subsection*{Acknowledgments}
The author would like to thank prof. Cestimir Burdik for warm hospitality in Prague.
 The paper is supported by Polish NCN grant  2014/13/B/ST2 and EU Cost Action MP1405 QSPACE.

\end{document}